\input{aipcheck}

\documentclass[
    ,final            
  ]{aipproc}

\layoutstyle{6x9}

\newcommand{\BEQ}{\begin{equation}}
\newcommand{\EEQ}{\end{equation}}
\newcommand{\BEA}{\begin{eqnarray}}
\newcommand{\EEA}{\end{eqnarray}}

\renewcommand{\d}{{\rm d}}

\newcommand{\re}{{\rm e}}

\newcommand{\si}{\hat{\sigma}}

\newcommand{\sip}{\hat{\sigma}_{+}}
\newcommand{\simin}{\hat{\sigma}_{-}}

\newcommand{\ha}{\hat{a}}
\newcommand{\half}{\frac{1}{2}}
\newcommand{\HH}{\hat{H}}
\newcommand{\HHS}{\hat{H}_{\rm S}}  
\newcommand{\HHW}{\hat{H}_{\rm W}}

\newcommand{\Hxi}{\hat{\xi}}

\newcommand{\HX}{\hat{X}}  
  
\newcommand{\HV}{\hat{H_I}} 
 
\newcommand{\RS}{{\rm S}} 
\newcommand{\RW}{{\rm W}} 
\newcommand{\RB}{{\rm B}}

\newcommand{\ri}{{\rm i}} 
\newcommand{\Is}{1} 
\newcommand{\Ig}{2} 
\begin{document} 
\title {Thomson's formulation of the second law:\\
an exact theorem and limits of its validity}

\author{Armen E. Allahverdyan}
{address={Yerevan Physics Institute,
Alikhanian Brothers St. 2, Yerevan 375036, Armenia }}

\author{Roger Balian}
{address={Service de Physique Th\'eorique, CEA Saclay,
91191 Gif-sur-Yvette cedex, France}}

\author{Theo M. Nieuwenhuizen}
{address={Institute for Theoretical Physics, 
Valckenierstraat 65, 1018 XE Amsterdam, The Netherlands}}

\begin{abstract}
  Thomson's formulation of the second law - no work can be extracted
  from a system coupled to a bath through a cyclic process - is
  believed to be a fundamental principle of nature.  For the
  equilibrium situation a simple proof is presented, valid for
  macroscopic sources of work.  Thomson's formulation gets limited
  when the source of work is mesoscopic, i.e. when its number of
  degrees of freedom is large but finite. Here work-extraction from a
  single equilibrium thermal bath is possible when its temperature is
  large enough. This result is illustrated by means of exactly
  solvable models.  Finally we consider the Clausius principle: heat
  goes from high to low temperature. A theorem and some simple
  consequences for this statement are pointed out.
\end{abstract}
\pacs{PACS: 03.65.Ta, 03.65.Yz, 05.30}

\maketitle

\subsection{\bf Introduction}
Since one century the laws of thermodynamics no longer appear as
basic principles, but as consequences of the laws of matter at the
microscopic level.  Their derivation from first principles requires an
identification of concepts such as temperature, work or heat in terms
of the elementary constituents, and an elimination of the irrelevant
microscopic variables by means of methods of statistical mechanics.
This program has been partially achieved for macroscopic systems in
the thermodynamic limit \cite{landau,balian}.
Here we focus on Thomson's (later Lord Kelvin)
formulation of the second law for equilibrium systems. 
We have two basic goals. The first is to discuss a rigorous version of this
formulation and to present a simple proof for it. This version uses a
minimal number of assumptions and has a larger region of applicability as
compared to the text-book presentation of Thomson's law
\cite{landau,balian}, where it is typically derived from the law of
entropy increase. Our second goal is to determine mesoscopic limits of
applicability of the formulation. They arise due to the existence of
large but not very large number of degrees of freedom (mesocopicity).
The special importance of Thomson's formulation comes from the fact
that other formulations rely more explicitly upon notions such as
entropy or temperature which require rather elaborate identification
in terms of microscopic variables, whereas this one concentrates on
the concept of work (energy exchange) which can be considered as a
primary, microscopic concept, as we show below.

\subsection{\bf General setup for Thomson's formulation}

In its original form Thomson's formulation states that 
if a thermodynamic system $\RS$, which interacts with a single 
macroscopic thermal bath
$\RB$ at uniform temperature $T$ and with a source of work $\RW$, 
evolves in a closed cycle, it cannot yield work
to the latter external system $\RW$. In other words, 
such a process cannot extract energy from the heat bath.

We assume that interaction of B
with S is sufficiently weak, yielding a negligible contribution
to the energy. (Later in this section we will comment on the situation
with strong coupling with the bath). Over sufficiently long times,
energy transfer identified with
heat may result from this interaction. The bath 
can therefore set up a canonical equilibrium at the temperature $T$ for 
S, without interacting directly with the source of work. 

In order to separate from each other the effects of the interactions
of S with B and with W, we shall consider in the following a cyclic
process that takes place in two steps. A canonical thermal equilibrium
of S is assumed to have already been established for times $t<0$.  The
interaction process that couples S and W takes place between the times
$t=0$ and $t=\tau$. The delay $\tau$ is supposed to be sufficiently
short so that no heat is exchanged with the bath during this first
step of the process. This is possible due to a weak coupling between
the system S and the bath B.  During the second step, for $t>\tau$,
the energy of W is left unchanged, while S returns to its original
thermal equilibrium state at temperature $T$ through interaction with
the bath B. Altogether, if the cycle is closed as regards the system
S, a positive amount of energy should according to Thomson have been
transferred from W to S between the times $t=0$ and $t=\tau$, and this
amount should be released from S to B for $t>\tau$. This two-step
analysis of the process will allow us to focus first on the coupled
evolution of S and W in the time-interval $0<t<\tau$, and to postpone
the forthcoming thermal exchange with the bath.  

More precisely, in order to define the {\it work} done by the source W, we
express the Hamiltonian which generates the evolution in this time-interval 
as
\BEA
\label{ham}
\HH=\HHS+\HV+\HHW,
\EEA
where $\HHS$ ($\HHW$) depends only on the variables of $\RS$ ($\RW$),
with $[\HHS,\HHW]=0$,
and where $\HV$ describes the coupling between $\RS$ and $\RW$. To fix 
ideas we shall first take $\HV$ in the form:
$\HV=\HX\Hxi$.
The observable $\HX$ belongs to the system $\RS$, and corresponds to a
``force'' variable such as the pressure for a gas or the magnetic
moment for a magnetic material. The conjugate variable $\Hxi$ refers
to the system $\RW$, and corresponds to a ``position'' variable such
as the volume for a gas or the magnetic field for a magnetic source.
Work done by S on W till a moment is identified as the energy yielded
by the source $\RW$, or equivalently as the decrease of the
expectation value $\HHW$ of its Hamiltonian \cite{balian}:
$W=\langle H_{\rm W}\rangle_0-\langle H_{\rm W}\rangle_t$.

\subsection{\bf Thomson's formulation for macroscopic sources}
In the thermodynamic limit for the system $\RW$ (but not necessarily for the 
system $\RS$), the fluctuations in $\Hxi$ are negligible at all relevant times;
in relative value they are small as the inverse square root of the number of 
microscopic degrees of freedom of $\RW$. The operator $\Hxi$ can therefore be 
replaced by its expectation value 
$\langle \Hxi\rangle_t={\rm tr}\,[\Hxi\rho_{\rm W}(t)]
\equiv \xi (t)$, $\rho_{\rm W}(t)={\rm tr}_{\rm S}\rho(t)$,
where $\rho(t)$ is the total density matrix of S and W, and where
$\rho_{\rm W}(t)$ is the marginal density matrix
of $\RW$. The evolution of $\xi (t)$ is generated by the Ehrenfest equation:
$\hbar\,\d\xi (t)/\d t=\ri\langle
\,[\HH,\Hxi]\,\rangle_t$.

 More precisely, if the fluctuation $\langle\Delta\Hxi^2\rangle_t$, where 
 $\Delta\Hxi=\Hxi-\langle\Hxi\rangle_t$, is negligible {\it at all times},
 the Schwartz inequality
 $|\,\langle\Delta\Hxi\,\Delta \hat{Z}\rangle_t\,|^2\le 
\langle\Delta\Hxi^2\rangle_t\,\langle\Delta \hat{Z}^2\rangle_t$,
where $\hat{Z}$ is any operator of S or W, implies that we can make the
replacement of the operator $\Hxi$ by its expectation value $\langle
\Hxi\rangle_t=\xi(t)$ everywhere, in particular in its Hamiltonian,
which thus takes the form
$\HH=\bar{H}_{\rm S}(t)+\HHW=\HHS+\HX\,\xi(t)+\HHW$,
where an effective time-dependent Hamiltonian
$\bar{H}_{\rm \RS}(t)$ appears for the system S.
This form of $\HH$ implies the factorization of the evolution operator,
and that of the density operator
$\rho(t)=\rho_{\RS}(t)\otimes \rho_{\RW}(t)$
at all times, provided the initial state at $t=0$ is also factorized.
Hence in the Schr\"odinger picture the density matrix $\rho_{\rm S}(t)$ 
of the system $\RS$ evolves according
to the unitary transformation $U$ generated by $\bar{H}_{\rm S}(t)$.

The equilibrium condition at time $t=0$ provides the initial density operator
of $\RS$ as
\BEA
\label{panama1}
\rho_{\rm \RS}(0)=\re^{-\bar{H}_{\rm \RS}(0)/T}/Z,
\qquad Z={\rm tr}\,\re^{-\bar{H}_{\rm \RS}(0)/T}.
\EEA
A cyclic transformation is characterized here by the fact that $\xi(t)$
has returned to its initial value at time $t=\tau$:
$\xi(\tau)=\xi(0),\qquad \bar{H}_{\rm S}(\tau)=\bar{H}_{\rm S}(0)$.
Owing to the lack of fluctuations in $\Hxi$, this condition is
sufficient to ensure that the overall process can be rigorously cyclic
at the microscopic level. Indeed, since the state of S is not
correlated at the time $t=\tau$ with that of W, it can return later on
to its original form (\ref{panama1}) at $t=0$ through
interaction with the thermal bath B.  The work given by $\RW$ to $\RS$
between the times $0$ and $\tau$ is equal due to: 
$W={\rm tr}\,[\,
\rho(\tau) \,\bar{H}_{\rm
S}(0)-\rho_{\rm S}(0)\,\bar{H}_{\rm S}(0)\,]$.  With help of
(\ref{panama1}) one gets, using the fact that $U$ is unitary:
\begin{eqnarray}
\label{rel}
\label{panama2}
\frac{W}{T}={\rm tr}\,[\, 
\rho_{\rm S}(\tau)\ln \rho_{\rm S}(\tau)-
\rho_{\rm S}(\tau)\ln\rho_{\rm S}(0)\,]\ge 0.
\end{eqnarray}
This expression is nothing but the {\it relative entropy}
of $\rho_{\rm S}(\tau)$ with respect to $\rho_{\rm S}(0)$;
it is strictly positive if $\rho_{\rm S}(\tau)\not=\rho_{\rm S}(0)$ 
\cite{sh}.

As shown by (\ref{panama2}), the work is positive for a macroscopic source 
of work W, whether the system $\RS$ is macroscopic 
or microscopic. Thus, in the second case this entails an extension of the 
original Thomson's statement~\cite{lenard,ANthomson}, which
was verified in specific examples~\cite{ANviola,ANNMR}. 
The irreversibility of the process arises
from the lack of symmetry between the times $0$ and $\tau$ which is
introduced by the initial condition (\ref{panama1}). Usually, the breaking
of time-reversal-invariance is associated with a dissipative process.
Here the latter process is the interaction with the heat bath, which will 
take place
after the time $\tau$. 

The above treatment can readily be generalized to an interaction with several
(not necessarily independent)
sources of work, where $\HV=\sum_{i=1}^r\HX_i\Hxi_i$.
Altogether, if the source $\RW$ of work is a macroscopic system, the
lack of fluctuations in the variables $\Hxi_i$, which couple $\RW$ to
the system $\RS$, ensures the second law in Thomson's form as soon as
$\RS$ is initially in a canonical equilibrium state. 

\subsection{ The fate of Thomson's formulation for mesoscopic sources.}
Let us now turn to the situation where $\RW$ is no longer a macroscopic 
object. We have to deal with the operator nature of the $\Hxi_i$'s 
and we can no longer leave aside the Hamiltonian $\HHW$
of the source of work, when studying the dynamics of $\RS$.

Work is still defined as the decrease of the energy of the source.  We
will again assume that in its initial state the system $\RS$ is in
canonical thermal equilibrium, and is not correlated with $\RW$:
$\rho(0)=\rho_{\RS}(0)\otimes\rho_{\RW}(0)$, where $\rho_{\RS}(0)$ is
given by (\ref{panama1}) with $\bar{H}_\RS(0)
=\HHS+\sum_{i=1}^r\HX_i\langle\Hxi_i\rangle_0$. A detailed discussion
on the necessity of this initial conditions is given in
\cite{ABNjcfin}.  However, the definition of a cyclic process
is no longer unambiguous. This
condition may indeed be generalized in two different ways, both of
which are rather natural \cite{ABNjcfin}.  To keep our presentation
concise we will here outline only one of the two definitions proposed
in \cite{ABNjcfin}. Since the variables $\HX_i$ and $\Hxi_i$ can now
be correlated, and since the interaction term $\HV$ 
involves a sum of their products, we search a
condition for such a sum of products.  We notice that in the
considered mesoscopic case the states of $\RS$ and $\RW$ do become
correlated for $t>0$.  Since now both $\rho(t)$ and $\HV$ irreducibly
live in the common Hilbert space of $\RS$ and $\RW$, the most standard
way to define energy exchange between two systems is to require that
the average interaction Hamiltonian does not contribute to the total
energy budget, i.e., it is zero both initially and finally.  For our
situation the proper interaction Hamiltonian is
$\sum_{i=1}^r\HX_i(\Hxi_i- \langle\Hxi_i\rangle_0)$, since its average
is zero at the initial time $t=0$. Requiring that this average is also
equal to zero at some final time $\tau$, we get for a cyclic process
with the duration $\tau$:
\BEA
\label{baba2}
\sum_{i=1}^r\langle\HX_i\Hxi_i\rangle_{\tau}=
\sum_{i=1}^r\langle\HX_i\rangle_{\tau}\langle\Hxi_i\rangle_0.
\EEA
For a large system $\RW$ we recover the macroscopic
condition for $r=1$, since the lack of fluctuations in $\Hxi$ implies
the lack of correlations with $\HX$:
$\langle\HX\Hxi\rangle_t\simeq\langle\HX\rangle_t
\langle\Hxi\rangle_t$.

{\it Work in the short-time limit.}
We shall discuss for (\ref{baba2}) the sign of the work integrated
over the duration $\tau$ of a closed cycle, in the limit where $\tau$
is sufficiently small so that we can expand $W$ in powers of $\tau$.
This limit is especially interesting, since mesoscopic contributions to
the work can overcome the main thermodynamic term: as we shown
\cite{ABNjcfin}, the thermodynamic contribution to the work, evaluated
by neglecting the fluctuations of $\Hxi_i$'s and given by
(\ref{panama2}), scales as $\tau^6$ for small $\tau$, whereas in the
mesoscopic case $W\sim\tau^2$.
Expanding the Heisenberg equations of motion one gets for the
work~\cite{ABNjcfin}:
\BEA
\label{halep}
\frac{2W_2}{\tau^2}&&=
\sum_{i,k=1}^r\langle ~\{\Delta\Hxi_i,\Delta\Hxi_k\}_+\,\rangle
\sum_{\alpha<\beta}(p_\alpha-p_\beta)
(h_\beta-h_\alpha)
\Re\{\,\langle \alpha |\HX_i|\beta\rangle
\langle \beta |\HX_k|\alpha\rangle\,\} \\
&&+\ri\sum_{i,k=1}^r\langle\,[\Hxi_k,\Hxi_i]\,\rangle 
\sum_{\alpha<\beta}(p_\alpha+p_\beta)
(h_\beta-h_\alpha)\Im\{\,\langle \alpha |\HX_k|\beta\rangle
\langle \beta |\HX_i|\alpha\rangle\,\},
\label{basra}
\EEA
where $\{...,...\}_+$ denotes anti-commutator, $\Re$ and $\Im$ stand
for the real and imaginary parts, respectively. $\{|\alpha\rangle\}$
is the common eigenbase of $\rho_\RS (0)$ and $\bar{H}_\RS (0)$, and
$p_\alpha$, $h_\alpha$ are the corresponding eigenvalues:
$p_\alpha=\re^{-\beta h_\alpha}/\sum_\alpha \re^{-\beta h_\alpha}$.
For $r=1$, where only the term (\ref{halep})
survives, one can show that $W>0$ because $\hat{\rho}_\RS
(0)$ is a decreasing function of $\hat{H}_\RS$.  Since
the thermodynamic contribution to the work is negligible
small, the amount of dissipated work predicted by the second
law, is {\it enhanced} in the mesoscopic domain.  The contribution
(\ref{basra}) reflects an effect of interference between different
types of work (channels), since it contains non-diagonal terms over
$i,k$ and becomes zero with $\langle\,[\Hxi_k,\Hxi_i]\,\rangle$.  Now
assume that the temperature of the system $\RS$ is so high
(fluctuations are strong) that $p_\alpha\simeq p_\beta$ for all the
$\beta$ and $\alpha$ that are connected by matrix elements of $\HX_k$
and $\HX_i$.  Then the contribution of (\ref{halep}) can be
disregarded, and provided that $\langle\,[\Hxi_k,\Hxi_i]\,\rangle\not
=0$ and $\Im\{\,\langle \alpha |\HX_k|\beta\rangle \langle \beta
|\HX_i|\alpha\rangle\,\}\not =0$, the sign of (\ref{basra}) can be
made negative.  Thus the many-source situation can lead to a violation
of the second law.  The above results also apply to the classical
limit, where Poisson brackets take the place of commutators.

Let us then examine the thermalization process which takes place for
$t>\tau$.  It is straightforward to show \label{ABNjcfin} that due
to the taken definition of cyclic processes, marginal density operator
of S is the equilibrium distribution (\ref{panama1}), both at the
preparation time $t=0$ and after the interaction with the thermal bath
that takes place for $t>\tau$.

{\it Exactly solvable models.}
The result above on the violation and enhancement of the Thomson's
formulation of the second law do not use specific properties of the
involved systems S and W, but unfortunately are essentially restricted
to short times. Moderate and long times can be explored by help of
exactly solvable models for S and W. To this end we investigated two
such models \cite{ABNjcfin,NBA}. The first of them is Jaynes-Cummings
model describing interaction of two-level atom or spin and a single
mode of radiation \cite{jaynes}. As for our
purposes, we can naturally identify the two-level atom as the system
$\RS$, and the electromagnetic field -  which is controlled externally
and admits a well-defined macroscopic limit - as the source $\RW$.
The Hamiltonian of the model reads
\BEA
\label{hamjc}
\HH_{\rm S}={\footnotesize\half}\,
\hbar\omega\si_z ,\quad
\HH_{\rm W}=\hbar\omega \ha^{\dagger}\ha,\quad
\HH_{\rm I}=\hbar g(\sip \ha+\simin\ha^{\dagger}), \EEA where
$\ha$ ($\ha^{\dagger}$) is the annihilation (creation)
operators of the field, $\si_x$, $\si_y$, $\si_z$,
$\si_{+,-}=(\si_x\pm i\si_y)/2$ are the Pauli matrices of the spin, and
$g>0$ is the coupling constant. The initial state of the spin is thermal with
temperature $T$ and that of the mode is described by a coherent state
$|\alpha\rangle$ with an amplitude $\alpha$.  With help of the known
exact solution of this model we found some new mechanisms of the
Thomson's law violation in the mesoscopic regime; for a detailed
discussion see \cite{ABNjcfin}. 

In the second exactly solvable model \cite{NBA} the systems S and W are two
ensembles of $N$ independent classical harmonic oscillators, coupled one-to-one
via their coordinates $x$ and $y$.
Like in an ideal gas, one may describe only one set of them, with Hamiltonians
\BEA
 {H}_{\rm S}&=&\half m\dot{ {x}}^2+\half m \omega^2  {x}^2,\quad
 {H}_{\rm W}=\half M\dot{y}^2+\half M\Omega^2 {y}^2, \quad
  {H_I}=-g\, {x} {y},
\EEA
The $x$-oscillator, associated with S, starts its evolution from a
thermal state at a temperature $T$. The $y$-oscillator starts from a
macroscopic state with a well-defined initial coordinate $y(0)$ and
velocity $\dot{y}(0)$. It plays the role of the work-source W. This
model not only allows to discuss limits of the second law for one
closed cycle, but actually admits rather detailed many-cycle
considerations \cite{NBA}. We determined conditions under which an
infinity of cycles with constant yield is possible.  This is a
mesoscopic 
`perpetuum mobile' that may cool the macroscopic bath B \cite{NBA}.

\subsection{ Clausius principle and its microscopic foundations.}  

Finally, we
shortly analyze the Clausius principle: heat goes from higher
temperature to lower one. To put this statement on the microscopic
ground, one considers two interacting systems 1 and 2 with the total
Hamiltonian \cite{landau,balian,tasaki,Reents}: $\hat{H}_{\rm
tot}=\hat{H}_\Is+\hat{H}_\Ig +\hat{H_I}$, where $\hat{H}_\Is$
($\hat{H}_\Ig $) is the Hamiltonian of the subsystem $\Is$ ($\Ig$),
and $\hat{H_I}$ is the interaction Hamiltonian.  The initial state
consists of two equilibrium subsystems at different temperatures,
$\rho_{\rm tot}(0)=\rho_\Is (0)\otimes\rho_\Ig (0)$, where
$\rho_{k}(0)=\exp[-H_{k}/T_{k}]/Z_{k}$ for $k=\Is\,,\Ig $.

Let us denote $ \Delta U_\Is=\langle \hat{H}_\Is\rangle_t-
\langle \hat{H}_\Is\rangle_0,$ 
$\Delta U_\Ig =\langle \hat{H}_\Ig \rangle_t-
\langle \hat{H}_\Ig \rangle_0$ and 
 $ \Delta U_I=U_I(t)-U_I(0^+)$.
It was rigorously proven that ~\cite{tasaki}:
\BEA
\frac{\Delta U_\Is}{T_\Is}
+\frac{\Delta U_\Ig }{T_\Ig }\ge 0,
\label{rashid}
\EEA
whereas correct but less rigorous arguments towards its validity are
provided in \cite{landau}. After the coupling between the
two subsystems has been turned on, the total energy
is conserved. Thus we get from (\ref{rashid}):
$(T_\Ig -T_\Is)~\Delta U_\Is\ge T_\Is\Delta U_{I}$.
In case ${T_\Ig >T_\Is}$  and $\Delta U_{I}\ge 0$
this inequality proves the Clausius principle.
If $\hat{H}_I$ commutes with $\hat{H}_1$ and $\hat{H}_2$, 
$U_I$ is conserved, implying that $(T_\Ig -T_\Is)~\Delta U_\Is\ge 0$,
as proven by Reents~\cite{Reents}.

When  ${T_\Ig >T_\Is}$  but $\Delta U_{I}<0$ there may, 
in principle, be a flow from low to high temperature, 
but it is bounded  if $\Delta U_{I}$ is bounded.
In particular, such a current cannot be constant and everlasting.

\end{document}